\documentclass{optica-article}

\journal{opticajournal} 

\articletype{Research Article}

\usepackage{lineno}
\usepackage{./styles/customcommands}

\begin{document}

\title{Mitigating tilt-induced artifacts in reflection ptychography via optimization of the tilt angles}

\author{Sander Senhorst,\authormark{1, *} Yifeng Shao\authormark{1}, Sven Weerdenburg\authormark{1,2}, Roland Horsten\authormark{1}, Christina Porter\authormark{2}, and Wim Coene\authormark{1,2}}

\address{\authormark{1} Department of Imaging Physics, Delft University of Technology, Lorentzweg 1, 2628 CJ Delft, The Netherlands\\
\authormark{2} ASML Netherlands B.V., De Run 6501, 5504 DR Veldhoven, The Netherlands\\
}

\address{\authormark{*}Corresponding author. Email:} \email{s.senhorst@tudelft.nl} 


\begin{abstract*} 
Ptychography in a reflection geometry shows great promise for non-destructive imaging of 3-dimensional nanostructures at the surface of a thick substrate. A major challenge to obtain high quality reflection-ptychographic images under near-grazing conditions has been to calibrate the incidence angle used to straighten the measured curved diffraction patterns in a process referred to as 'tilted plane correction' (TPC). In this work, we leverage the flexibility of automatic differentiation (AD)-based modeling to realise an alternative approach, where the tilted propagation is included into the forward model. Use of AD allows us to jointly optimize the tilt angles with the typical probe and object, eliminating the need for accurate calibration or random search optimization. The approach was validated using datasets generated with an extreme ultraviolet (EUV) beamline based on either a tabletop high harmonic generation (HHG) source or a visible laser. We demonstrate that the proposed approach can converge to a precision of $\pm 0.05^\circ$ for probe beams at $70^\circ$ angle of incidence, possibly precise enough for use as a calibration approach. Furthermore, we demonstrate that optimizing for the tilt angles reduces artifacts and increases reconstruction fidelity. Use of AD not only streamlines the current ptychographic reconstruction process, but should also enable optimization of more complex models in other domains, which will undoubtedly be essential for future advancements in computational imaging.
\end{abstract*}

\section{Introduction}
Ptychography is a form of coherent diffraction imaging (CDI) where a coherent probe beam is scanned across a sample while the resulting diffraction pattern is measured at each scanning position. Because neighbouring probe positions overlap, the redundancy in the diffracted intensity between adjacent scanning positions over-constrains the ill-posed inverse problem of retrieving the complex field of the object and probe. In this way, ptychography moves the complexity of imaging from the instrumentation side to the computational domain. Since ptychography can produce high-resolution images without the need of diffraction-limited illumination or objective optics, it has seen considerable use wherever availability of high-quality optics is problematic, such as in the extreme ultraviolet (EUV) wavelength regime\cite{sandberg_lensless_2007, gardner_high_2012,lu_characterisation_2023, esashi_tabletop_2023,odstrcil_towards_2019, eschen_material-specific_2022}.

To make use of ptychography for imaging samples on thick substrates, one must make use of a reflection geometry, where the detector is placed near the direction of specular reflection for the substrate surface. Most materials in the EUV regime are only weak reflectors, so obtaining sufficient diffracted intensity is a challenge. Since shallow incidences increase the reflected intensity, the incidence angles for EUV reflection measurements are usually set close to grazing, generally between 70$^\circ$ and 80$^\circ$ with respect to the surface normal. The disadvantage of large incidence angles is the complication of the reconstruction process, as the tilt between the sample and detector planes distorts the diffraction patterns from a rectilinear to a curved grid\cite{patorski_fraunhofer_1983}. 

Conventionally the diffraction pattern curvature is corrected by interpolating the collected diffraction patterns in a pre-processing step before feeding the diffraction data to the reconstruction algorithm in a process called tilted-plane correction (TPC)\cite{marathe_coherent_2010, gardner_high_2012, porter_general-purpose_2017, seaberg_tabletop_2014, lu_characterisation_2023}. Application of TPC requires the experimental geometry (i.e. propagation distance and rotation angles) to be well-known, although most published work does not specify how these parameters were obtained. This is relevant, since we shall see that already a relatively small error of $2^\circ$ is sufficient to cause significant artifacts in the reconstructions. In practice, the required parameters for TPC may be obtained either through precise calibration of the experimental geometry, manual estimation through symmetry in the diffraction pattern\cite{porter_complex_2019} or simple trial and error. Experimental calibration, although preferable, causes significant experimental challenges and requires re-calibration after every system adjustment. Usage of diffraction pattern symmetry assumes the sample to be real-valued, which is only true for simple samples without height variation. Additionally, it assumes the sample diffraction orders to be sufficiently discernible with respect to the numerical aperture (NA) of the incident probe beam, which will not hold as incident NA increases. Finally, trial and error is a pain-staking process involving many sub-optimal reconstructions, which may introduce selection biases by the observer into the final reconstructions. In contrast, algorithmic correction of the uncertain tilt angle has the benefit of objectivity, as biases due to manual estimation are removed. In addition, it may offer extra robustness, since it places no additional constraints on the sample and illumination.

Recently an algorithm has been proposed which applied TPC as part of the reconstruction process, by the name of aPIE \cite{beurs_apie_2022}. Using a random search-type approach, the authors demonstrated convergence and quality improvement for illumination at $45^\circ$ angle of incidence in the visible regime, however the precision was significantly lower for general non-structured beams. Moreover, the relatively small tilt and detection NA contributed to a less curved coordinate transformation than is generally the case for experimental data measured in high tilt EUV reflection ptychography. 

In this work we will attempt to extend the range of applicability of tilt angle optimization to EUV ptychography, with the goal of achieving accurate convergence for large tilt angles and general experimental illumination conditions, not just structured light. Note that, although diffusers for generating structured illumination are emerging in the EUV regime\cite{eschen_structured_2024}, they are not yet widespread and still increase measurement time due to absorption. Furthermore, in applications where ptychography may be used as a calibration approach for EUV metrology systems\cite{du_high-resolution_2023} like grating profilometry in wafer metrology, structured illumination is generally not an option since the goal is often to characterize a nearly-Gaussian beam. 

To achieve accurate convergence we take an alternative approach compared to aPIE: we leverage the flexibility of our automatic differentiation (AD)\cite{kandel_using_2019}-based ptychographic reconstruction framework \cite{shao_wavelength-multiplexed_2024} to jointly optimize for the tilt angles together with the other parameters of the forward model. In contrast to traditional ptychographic iterative engine (PIE) methods\cite{rodenburg_phase_2004, maiden_improved_2009, maiden_further_2017, maiden_wasp_2024}, which require defining an update step explicitly for every model parameter, using AD we can obtain the gradient with respect to all model parameters, including the tilt angles, simultaneously. This means that operations whose gradients are non-trivial to analytically differentiate, such as the interpolation operation that will be used in our tilted forward model, are now available in a fully differentiable manner. This makes the AD approach particularly versatile in extending the ptychographic model to include experimental uncertainties, as long as the model that describes these uncertainties is numerically differentiable.

\medskip

\section{Materials and Methods}
In our reflection-type experiment, the 2D planes of the pupil, sample and detector are mutually non-parallel, requiring computation of diffraction between tilted planes. We will need to model the light propagation between these planes to predict the diffraction patterns from our ptychographic datasets. To achieve this we use the fact that the direction of the different plane wave components $\vec{k}$ of scalar fields are preserved upon propagation. Computation of light propagation between tilted planes is then reduced to application of the adequate coordinate transformation to yield the plane waves in the different sample and detector coordinate planes.

\begin{figure}[!htbp]
    \centering\includegraphics[width=5.5in]{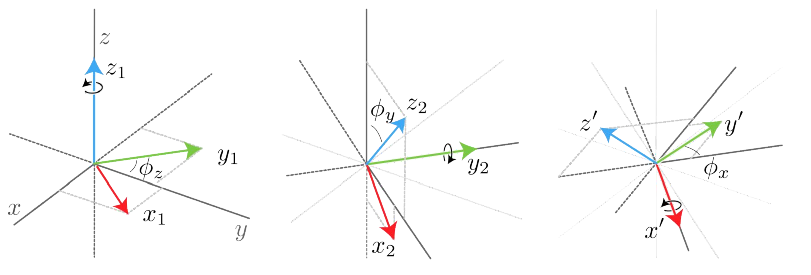}
    \caption{The definition of the rotation angles $\phi_z, \phi_y, \phi_x$, including the intermediate co-rotating coordinate systems. From left to right, first the non-primed coordinate system is rotated by $\phi_z$ about the $z$-axis, giving the $(x_1, y_1, z_1)$ coordinates. Then these coordinates are rotated by $\phi_y$ about the $y_1$ axis to give the $(x_2, y_2, z_2)$ coordinates. Finally, we rotate by $\phi_x$ about $x_2$ to give the final primed coordinates $(x', y', z')$.}
    \label{fig:rotation_transform}
\end{figure}

We will transform between the Cartesian coordinate systems of these planes through a three-dimensional rotation matrix $R$. There are multiple choices for defining 3D rotations; in this work we choose the rotation angles from the non-primed sample plane to the primed detector frame to correspond to intrinsic rotations of $\phi_z, \phi_y, \phi_x$ successively about the $z$, $y_1$, $x_2$ axes, where the subscript indicates the coordinate frame is co-moving with the transformation, as shown in Fig. \ref{fig:rotation_transform}. The rotation matrix describing this transformation is given by
\begin{align}
\begin{split}
    R(\phi_x, \phi_y, \phi_z) = R_x(\phi_x)R_y(\phi_y)R_z(\phi_z) = \begin{bmatrix}
R_{00} & R_{01} & R_{02}\\
R_{10} & R_{11} & R_{12}\\
R_{20} & R_{21} & R_{22}
\end{bmatrix}&\\
    =\left[\begin{smallmatrix}
1 & 0 & 0\\
0 & \cos\phi_x & \sin\phi_x\\
0 & -\sin\phi_x & \cos\phi_x
\end{smallmatrix}\right]\left[\begin{smallmatrix}
 \cos\phi_y & 0 & \sin\phi_y\\
 0 & 1 & 0 \\
-\sin\phi_y & 0 & \cos\phi_y
\end{smallmatrix}\right]\left[\begin{smallmatrix}
\cos\phi_z & \sin\phi_z & 0\\
-\sin\phi_z & \cos\phi_z & 0\\
0 & 0 & 1
\end{smallmatrix}\right]&.\\
\label{eq:rot_matrix_def}
\end{split}
\end{align}

Since for sufficiently thin samples it is possible for the sample to be modeled by a 2D complex reflection function, similar to the traditional ptychographic 2D complex transmission function, we model the probe $P$ and object $O$ as 2D functions in the non-primed sample coordinates. We also add a phase factor to the probe to account for the oblique incidence of the illumination. The exit field is then given by
\begin{equation}
    U_e(x, y) = P(x,y) O(x, y) \exp{ikx\sin \phi_{i,y}},
    \label{eq:probe_times_object}
\end{equation}
propagating in the positive z-direction, with the wavenumber $k=\frac{2\pi}{\lambda}$ and $\phi_{i,y}$ the angle of incidence with respect to the surface normal. Numerically, the addition of the exponential factor is equivalent to defining the origin of the detector coordinate plane in the direction of specular reflection. When interpreting the ptychographic reconstructions, one should however not forget that the Fourier transform of the exit field is always relative to this shift. This means that, to obtain the physical exit field, one must take care to include the exponential phase ramp.

\begin{figure}[!htbp]
    \centering\includegraphics[width=6in]{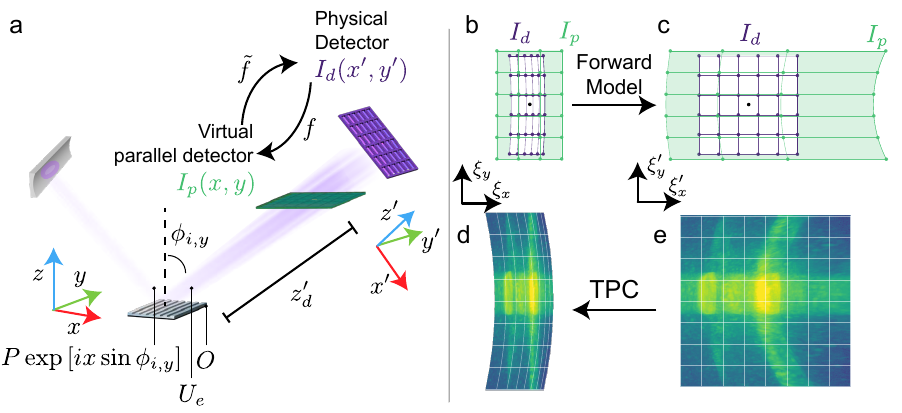}
    \caption{a) The different Cartesian coordinate systems, fields and intensities used in the forward model computations, for the case of a single non-zero tilt angle $\phi_y$. From the left, a coherent probe $P$ is incident on a diffracting sample $O$. After reflection by the sample, the exit field $U_e$ propagates in a rotated coordinate system for a distance $z'_d$, where a virtual detector in the non-rotated coordinates measures the parallel diffraction pattern $I_p$. $I_d$ is then given by interpolating $I_p$ onto the rotated far-field coordinates. b,c) In our new forward model approach, tilted diffraction patterns are predicted by interpolating the pixels sampled by $I_p$ onto the rectangular coordinates of $I_d$. Coordinates in the parallel/non-primed (b) transform to coordinates in the rotated/primed (c) frames using $\tilde{f}$. The coordinates for $I_p$ are rectangular in the parallel frame, while the coordinates for $I_d$ are rectangular in the rotated frame. The dot in the center indicates the origin. The shaded area can be considered as super-resolution information, since it contains pixels that are present in the forward model, but are never measured by the detector. In our forward model tilting approach, the pixels sampled by $I_p$ would be interpolated onto the rectangular coordinates of $I_d$. d,e) The traditional tilted plane correction approach. A measured diffraction pattern (e) $I_d$ is interpolated onto rectangular coordinates in the non-primed frame (d). As can be seen, the curved line corresponding to the $\xi_y'$-axis has been straightened due to the coordinate transformation.}
    \label{fig:coordinate_systems}
\end{figure}

To propagate the exit field $U_e$ to the detector, we assume that the detector is placed in the far-field, such that Fraunhofer propagation may be applied. In the far-field, the field at each measured position originates from a single plane-wave component of the exit field $U_e$; the direction of this plane-wave corresponds to the direction from the sample to the measured position. Since the direction of these plane waves are conserved upon propagation, a plane-wave decomposition of the exit field (i.e. the Fraunhofer propagation integral), should still suffice to compute the diffraction between tilted planes, so long as the transformation between the coordinate systems is incorporated\cite{wyrowski_fast_2003}. The relation between the field in the sample plane $U_e(x,y)$ and the detected intensity at the detector $I_d(x' ,y') $ is obtained in two steps. First, we compute the far-field intensity $I_p$ that would be measured by a detector which is placed parallel to the sample plane. This may be computed using the regular Fraunhofer propagation integral and taking the square modulus,
\begin{equation}
    I_p\left(\ksi_\bot\right) = \bigg \lvert \iint U_e(x, y) \exp{i(x \xi_x + y \xi_y)}dxdy \bigg \rvert^2,
    \label{eq:propagator}
\end{equation}
where $\ksi_{\bot}$ are the x- and y Fourier coordinates in the sample plane. In the second step we account for the change in coordinate systems due to rotation. Since we are in the far-field, this only involves introducing the rotated, primed coordinates in the argument of $I_p$, to obtain the actual measured intensity at the detector
\begin{equation}
    I_d\left(x', y'\right) = I_p(\ksi_\bot)\bigg\rvert_{\ksi_{\bot}=f\left(\ksi'_\bot\right)},
    \label{eq:tilted_propagator}
\end{equation}
where $\ksi_{\bot}' =\left(\tfrac{kx'}{z'_d}, \tfrac{ky'}{z'_d}\right)$ are the x and y Fourier coordinates in the detector plane, $k=\frac{2\pi}{\lambda}$ is the wavenumber for source wavelength $\lambda$, $z'_d$ is the the propagation distance from the sample to the detector and $f$ is the (non-linear) 2-dimensional coordinate transformation arising from the tilt, which will be subsequently derived. A schematic overview of the coordinate systems, fields and intensities is shown in Fig. \ref{fig:coordinate_systems}a.

When all three spatial dimensions are considered, the equivalent transform to $f$ is simply given by the 3D rotation $\ksi = R\ksi'$, with $R$ being a general rotation matrix. 
From this, we can extract the (x,y)-coordinates to obtain the two-dimensional transformation
\begin{equation}
    \ksi_\bot = f(\ksi'_\bot)= \begin{bmatrix}
        R_{00} & R_{01} & R_{02}\\
        R_{10} & R_{11} & R_{12}
    \end{bmatrix}\begin{bmatrix}
        \xi'_x\\
        \xi'_y\\
        \sqrt{k^2-\xi^{'2}_x - \xi^{'2}_y}
    \end{bmatrix},
    \label{eq:tilt_correction_transform}
\end{equation}
where we used the Ewald sphere criterion $|\ksi'| = k$ to find the $\xi'_z$-component. It should be noted that this transform is still linear with respect to $k$, so its non-linearity is only dependent on the numerical aperture ($\text{NA}$) of the spatial frequencies present in the transform, not the wavelength. The action of the coordinate transformation $f$ is shown in Fig. \ref{fig:coordinate_systems}b-c. In reflection ptychography, the experimental geometry is such that we expect rotation around just the y-axis with other tilt angles equal to zero. In this case, the origin of detector is transformed from $(0, 0)$ to $(k\sin\phi_y, 0)$. If we then assume $\phi_y = \phi_{i,y}$, this shift in the origin exactly cancels the shift introduced in Eq. \ref{eq:probe_times_object}, thus making the exit field equivalent to that used in transmission ptychography.

Although the relations in Eq. (\ref{eq:propagator}) and Eq. (\ref{eq:tilted_propagator}) are Fourier transforms, they cannot be computed using computationally favorable methods like the Fast Fourier Transform (FFT). The FFT requires a uniform sampling along each axis, while the function  $f$ to transform the coordinates only produces uniformly sampled coordinates in the limit of either small numerical apertures ($\ksi_z' \approx k$, $\phi_z = 0$) or small rotation angles ($\sin \phi \approx 0$). This is why conventional approaches apply TPC to interpolate the measured diffraction patterns $I_d$ onto a uniform grid in $\ksi_\bot$. In this work we will take a different approach by including the coordinate transformation $f$ into the forward model and interpolating the intensity detected by a virtual camera parallel to the sample $I_p$ onto the coordinates of the tilted camera to give the actual detected intensity $I_d$. In this case, the coordinate transformation in Eq. (\ref{eq:tilt_correction_transform}) should be inverted, which is again done by using the Ewald Sphere criterion to eliminate the $\xi_z$ coordinate:
\begin{equation}
    \ksi'_\bot = \tilde{f}(\ksi_{\bot}) \begin{bmatrix}
        (R^{-1})_{00} & (R^{-1})_{01} & (R^{-1})_{02}\\
        (R^{-1})_{10} & (R^{-1})_{11} & (R^{-1})_{12}
    \end{bmatrix}\begin{bmatrix}
        \xi_x\\
        \xi_y\\
        \sqrt{k^2-\xi^{2}_x - \xi^{2}_y}
    \end{bmatrix},
    \label{eq:tilt_forward_transform}
\end{equation}
where the $(R^{-1})_{ij}$ are components of $R^{-1}$. Since $R$ is a rotation matrix, we may simply use $R^{-1} = R^T$. The action of the coordinate transformation $\tilde{f}$ on rectangular coordinates is shown in Fig. \ref{fig:coordinate_systems}b-c. When computing $\xi_z$ one must take care to include the phase shift introduced in Eq. (\ref{eq:probe_times_object}) as a shift in the sampled coordinates. 

For forward modeling, we can first compute the diffraction pattern in a virtual parallel camera $I_p$, which is then interpolated onto the actual tilted camera $I_d$ using Eq. (\ref{eq:tilted_propagator}) and Eq. (\ref{eq:tilt_forward_transform}) through a bilinear interpolation, where it is important to notice that although $\ksi$ is uniform, $\tilde{f}(\ksi)$ is not. We then interpolate the output of $\tilde{f}(\ksi)$ to the uniform coordinate $\ksi'$, noting that both $f(\ksi)$ and $\ksi'$ are defined on the tilted camera plane. Since the result of bilinear interpolation is computed using a weighted average of the surrounding pixels, this result is differentiable with respect to the position of the transformed coordinates $\tilde{f}(\ksi)$, which are in turn a function of the rotation angles $\phi_x, \phi_y$ and $\phi_z$.

The loss function used for the optimization was equivalent to the loss previously introduced in\cite{shao_wavelength-multiplexed_2024}, with only a single additional term $\frac{\sum \mathcal{M}_i I_{p,i}}{\sum \mathcal{M}_i}$ added to the total loss. Here, the index $i$ refers to the flattened pixels of the $M$ by $N$ detector grid and $\mathcal{M}_i \in \{0, 1\}^{N M}$ is a mask which is 1 for pixels of the virtual parallel detector which fall outside the measurement domain of the physical detector and 0 for pixels inside. The pixels which fall outside the measurement domain correspond to the shaded area in Fig. \ref{fig:coordinate_systems}c. This additional term is required because every choice of sampling which preserves the resolution must have some pixels which are never interpolated onto the physical detector. If these pixels were not suppressed, this would allow the model degrees of freedom which would never be part of the loss function, increasing the risk of over-fitting. This term is still required if the sampling is chosen such that initially all pixels would be interpolated onto the detector; It is possible that pixels that initially fell onto the detector no longer do so as the tilt angles differ from their initial values during optimization.

The model is optimized using stochastic gradient descent (SGD). As such, we batch the total dataset, adopting the nomenclature from the machine learning community, where an iteration over the full dataset is called an epoch. If the total number of scanning positions is $K$, then we have an $\text{epoch}= \left \lceil{K/B}\right \rceil$ iterations, with $B$ the batch size. In this work, all optimizations use $B=1$. For more details on the optimization procedure and applied regularizations, we refer to the supplemental material.

Although we have focused on the optimization of the probe $P_i$, the object $O_i$ alongside the newly introduced rotation angles $\phi_x, \phi_y$ and $\phi_z$, it should be noted that the addition of tilting into the forward model is fully compatible with all optimizable variables introduced in earlier versions of the model\cite{shao_wavelength-multiplexed_2024}, such as wavelength, scanning position correction and propagation distance. This new approach thus allows for joint optimization of all model parameters simultaneously, so long as they are not redundant. In this work, we wish to assess specifically the performance of tilt angle optimization while avoiding possible interference due to optimization of other variables. As such, other experimental variables were fixed during all reconstructions.

\medskip
\section{Results \& Discussion}
To validate that the proposed algorithm is well-behaved and could generate reliable results by optimizing for the tilt angles, reconstructions were initially performed using experimental data in the visible wavelength regime. In addition, we will also report on artifact reduction and fidelity improvement for EUV datasets in sections \ref{sec:EUV_artifact} and \ref{sec:EUV_fidelity}.

\begin{figure}[htbp]
    \centering\includegraphics[width=\textwidth]{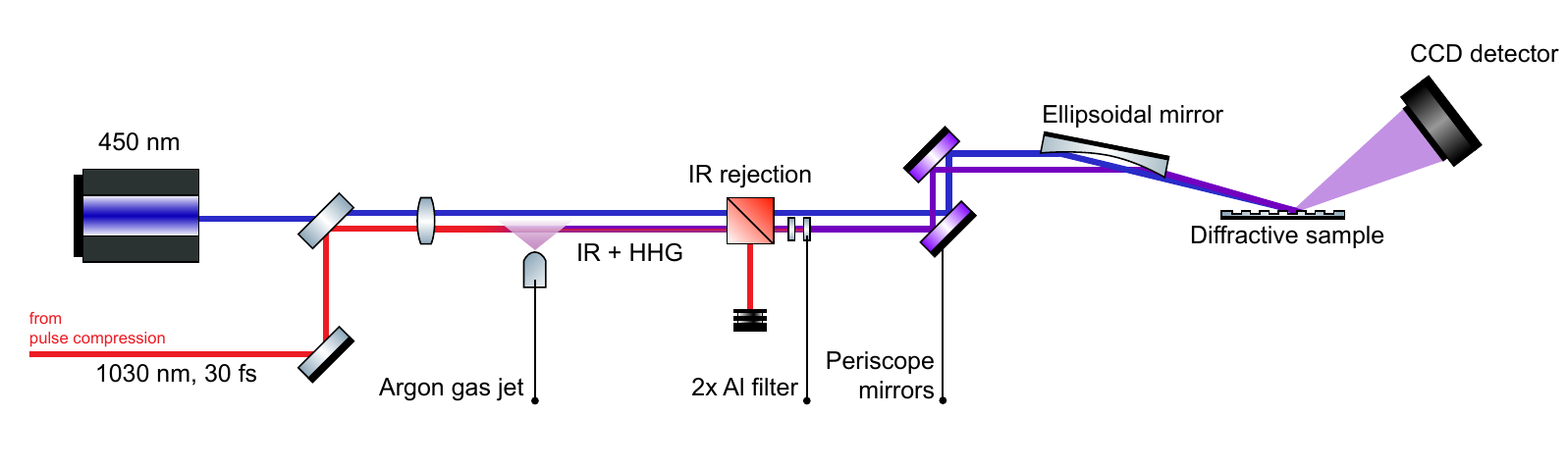}
    \caption{A schematic of the experimental setup used for ptychographic imaging. From the left, the input light consisted of either a 450 nm blue laser, or high-power pulsed infrared (IR) laser. The light is then focused close to a Argon gas jet to produce high harmonic generation (HHG) extreme ultraviolet (EUV) radiation. For the visible light experiments, the gas jet was disabled. The IR beam was subsequently removed from the beam path using IR-rejecting mirrors and pair of aluminum filters. For the $450$ nm illumination, the aluminum filters were removed. A periscope consisting of two multilayer mirrors select the 17.5 nm wavelength from the generated HHG spectrum, while the $450$ nm illumination is simply reflected. An ellipsoidal mirror then focuses the light onto the sample. The reflected light is finally measured by a CCD detector. For visibility the two beams in the figure have been separated while in the actual experiment they were co-linear.}
    \label{fig:optical_system_schematic}
\end{figure}

For the visible experiments, the illumination consisted of a $450$ nm laser loosely focused onto the sample using an ellipsoidal mirror designed for use in the extreme ultraviolet (EUV) wavelength regime, yet sufficiently reflective for visible wavelengths, providing an illumination NA of ca. $40$ mrad x $80$ mrad. The sample consisted a 1997 Toshiba computer chip, which was scanned through 970 scanning positions to cover an area of 1 mm$^2$ with an overlap of $R/2$, where $R$ is the approximate radius of the probe. The relatively large amount of scanning positions used was a result the desired field-of-view and is not expected to significantly impact the algorithm performance. A schematic of the optical system is shown in Fig. \ref{fig:optical_system_schematic}.

The incidence angle was set to approximately $\phi_y = 70^\circ \pm 1^\circ$, a realistic value for incidence angles used in the EUV regime. The detector geometry was such that x and z rotation angles are approximately zero. At each scanning position, the intensity of the diffraction pattern $I_d$ was measured at a distance $z'_d = 70$ mm, yielding a detection NA of approximately $0.3$. To reconstruct the datasets, the initial model probe was set to a general ellipsoidal aperture function with a uniform phase and the sample was set to a uniform reflector. 

In contrast to standard transmission ptychography approaches, the final interpolation step decouples the sampling between the primed and non-primed coordinates, allowing for a freedom of choice in sampling. Due to the rotation, the oversampling of the detector space with respect to the parallel detector has now become spatially dependent\cite{porter_general-purpose_2017}, but the ptychographic dataset has been shown to be able to sufficiently over-constrain the problem to properly reconstruct even in the case of sparse sampling\cite{edo_sampling_2013}, allowing freedom of choice in the sampling of the parallel detector space, and by consequence probe and object space. The sampling of the model was chosen with sufficient margin to ensure every detector pixel was interpolated onto, including a small margin to account for changes in the tilt angles. More details on sampling choices can be found in the supplemental material.

\subsection{Convergence}

The initial tilt angle $\phi_y$ was scanned through a range between $68^\circ$ and $72^\circ$ in steps of $0.5 ^\circ$, while $\phi_x$ and $\phi_z$ were initially set to 0. The optimization was performed using the Adam optimizer\cite{kingma_adam_2017} with a learning rate of $0.5$ for the probe and object and a learning rate of $0.005$ for the three tilt angles. All learning rates decayed exponentially to $1 \%$ of the original value at the end of the reconstruction. To prevent algorithm instabilities due to the large gradient for early iterations, the reconstruction was first performed on a fixed tilt angle for 10 epochs, after which the optimization for the tilt was optionally turned on. To further stabilise the reconstructions, the tilt angles were stored internally as $\phi=\arctan(a)$, where the parameter $a$ was optimized. This makes the training behavior for the angles $\phi$ asymptotic towards $\pm 90^\circ$, which prevents accidental overshooting due to large update step-sizes. To assess the algorithm performance, a comparison was made between reconstructions where the tilt angle was kept constant and identical reconstructions where the tilt angle was jointly optimized with the probe and object.

\begin{figure}[htbp]
    \centering\includegraphics[width=7in]{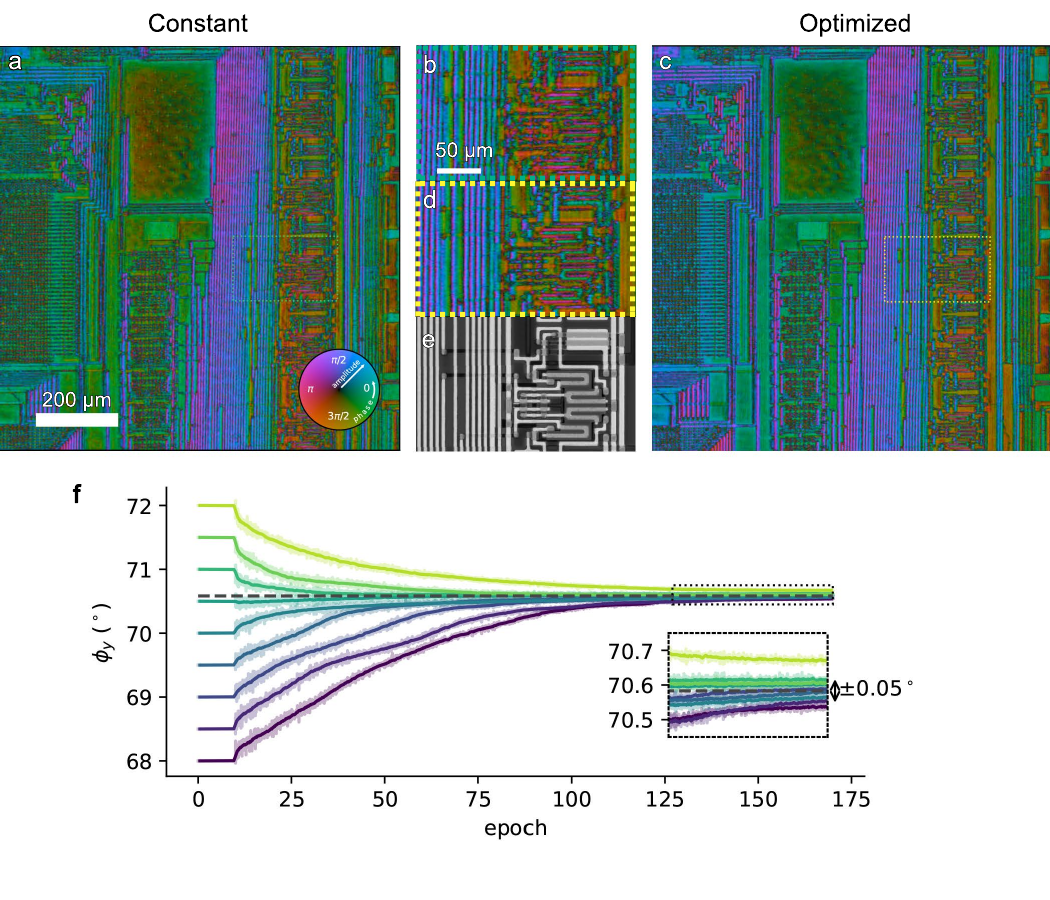}
    \caption{a,c) The reconstructed complex amplitude of the sample that falls completely within the range of the scanning positions after optimizing for 170 epochs using an initial angle of 68$^\circ$ without (a) and with (c) turning on tilt angle optimization. b,d) A region of interest of the constant (b) versus the optimized (d) reconstruction demonstrating visibly improved line sharpness and removal of ghosting effects. Phase contrast was enhanced by reducing the brightness of low-amplitude signals, as is reflected by the colormap. (e): a reference (non-complex) image taken by white light interferometry of the same region of interest. The amplitude contrast shows height difference rather than reflection amplitude, where black corresponds to $z=0$ and white corresponds to $z=2.5~\mu$m. (f) The optimization progression of $\phi_y$ for initial guesses of the tilt angle between 68$^{\circ}$ and 78$^{\circ}$ degrees with 0.5$^{\circ}$ increments. The first 10 epochs show a constant $\phi_y$, since only after 10 epochs the optimization was enabled. The inset shows the precision of the convergence by scaling the y-axis.}
    \label{fig:EX0085_convergence}
\end{figure}

The results of the ptychographic reconstructions for the part of the sample which satisfied the overlap condition are shown in Fig. \ref{fig:EX0085_convergence}. The entire field of view, including reconstruction artifacts at the edges, is visible in the supplemental material. For the visible data, the algorithm reliably converged to $\phi_y=70.58^\circ ~ \pm 0.05^\circ$. In comparison, the optimized values for $\phi_x$ and $\phi_z$ deviated by no more than $0.5^\circ$ from their original starting value of 0. The optimization progression for $\phi_x$ and $\phi_z$ can be found in the supplemental material. The high precision of the reconstruction for the crucial $\phi_y$ angle indicates this reconstruction technique may even be used as a calibration tool, where not the reconstructed probe and object are of interest, but rather the experimental geometry itself.

Interestingly, the optimization for initial angles which were too large compared to the convergence value progressed more slowly than for initial angles which were too small. We suspect this asymmetry is caused by the differences in sampling, since the initial guess for the tilt angles also determines the sampling to be used in the reconstructions.

\subsection{Artifact reduction}
\label{sec:EUV_artifact}
\begin{figure}[htbp]
    \begin{centering}\includegraphics[width=6in]{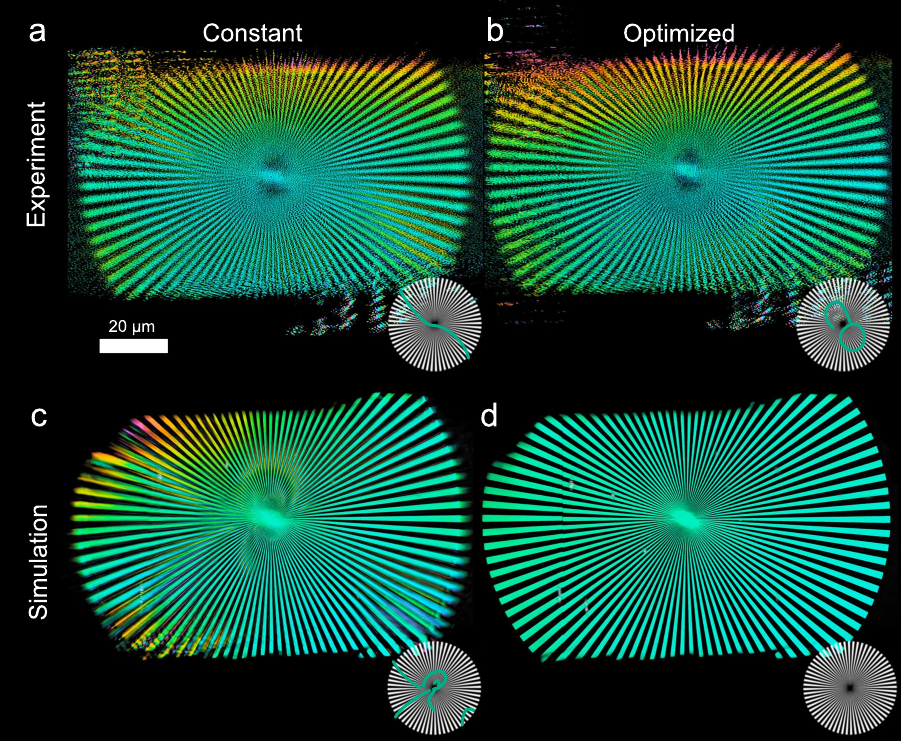}
    \caption{A reconstruction of the Siemens star from experimental data (a, b) and simulation data (c, d) for an initial angle $\phi_{y,\text{initial}}$ of 68$^\circ$ where the angle was kept constant (a, c) or optimized in the model (b, d). The inset indicates the areas of the Siemens star where the inner spokes are 'out of phase' with the outer spokes, demonstrating internal inconsistency. Since the reference was nearly identical to (d) it was not shown separately. Note to digital readers: the Siemens star is particularly prone to aliasing artifacts, so zoom accordingly.}
    \label{fig:EX0033_SI0007_siemens_star_ROI_1}
    \end{centering}
\end{figure}

EUV illumination took place in the same experimental setup as used for the visible datasets, where a 1030 nm Ytterbium-doped fiber laser (Active Fiber System UFFL 100, $100$ W average output power) was loosely focused into an argon gas jet to produce high-harmonic EUV radiation, out of which the $17.5$ nm wavelength harmonic was selected using a pair of wavelength-selective multilayer mirrors. This single harmonic was focused onto the sample using the same ellipsoidal mirror as for the visible light. The sample consisted of a 20 nm height gold Siemens star patterned on a silicon substrate, which was scanned over 225 scanning positions covering a range of $108 ~\mu\text{m}$ x $40~\mu\text{m}$ with an overlap of $R/3$, where $R$ is the probe radius in that direction. For more details on the experimental system, we refer to\cite{shao_wavelength-multiplexed_2024}. Reconstructions were performed using general initial guesses for the probe and object. For the probe, we made use of two coherence modes\cite{thibault_reconstructing_2013} to account for illumination partial coherence. The region of interest of the resulting reconstructions has been shown in Fig. \ref{fig:EX0033_SI0007_siemens_star_ROI_1}. The full reconstructed sample and probe are shown the supplemental material.

To verify the origin of the observed artifacts as arising from errors in the tilt angles, as opposed to other experimental errors, a simulation dataset was generated in addition to the visible and EUV datasets. The simulation dataset was designed to mimic the actual EUV experiment in the absence of any experimental limitations such as detector dynamic range, noise or scanning position errors. The simulation had ground-truth tilt angles of $(\phi_x, \phi_y, \phi_z) = (0^\circ, 70^\circ, 0^\circ)$, the scanning positions used were identical to the experimental ones. The scanning positions were coarsely aligned to illuminate the same part of the sample, which was generated based on a scanning electron microscope (SEM) reference image. The simulated sample complex reflectance was calculated by segmenting the SEM image into gold and silicon pixels and accordingly setting the Fresnel coefficient of reflection at the ground truth incident angle of 70$^\circ$. The probe used for simulation was retrieved from a previous reconstruction of the experimental dataset after manual removal of artifacts.

The Siemens star sample is particularly suitable for artifact analysis as its spokes are contiguous across a large field of view while requiring increasingly more placement precision towards the center. Thus, wherever the reconstruction is inconsistent between low and high spatial frequencies this generates a continuity artifact, where the inner spokes' position is exactly one spoke 'out of phase' with the outer spokes. As can be seen in Fig. \ref{fig:EX0033_SI0007_siemens_star_ROI_1}c, these artifacts appeared very prominently when reconstructing using a fixed $\phi_y$ of $68^\circ$, a deviation of only $2^\circ$ with respect to the ground truth of $70^\circ$. These artifacts completely disappeared when tilt angle optimization was enabled, which gave a final value which only deviated from the ground truth by $0.005^\circ$ for the simulation dataset. The corresponding experimental reconstructions (Fig. \ref{fig:EX0033_SI0007_siemens_star_ROI_1}a) with the same estimated tilt error also show the continuity artifacts going outward all the way to the edge of the Siemens star, while for the tilt optimized case the artifacts are less prominent and only appear below roughly half of the radius of the Siemens star, thus indicating the reconstructed sample is internally consistent (i.e. its lines are continuous) up to a higher spatial frequency.

\subsection{Fidelity improvement}
\label{sec:EUV_fidelity}
\begin{figure}[htbp]
    \begin{centering}\includegraphics[width=6in]{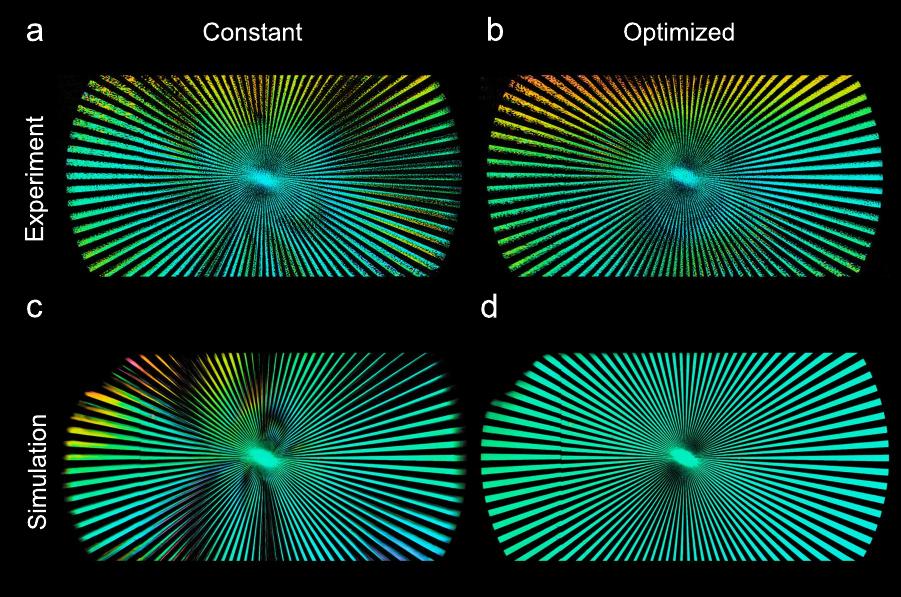}
    \caption{A multiplication of the absolute value of the SEM reference with the reconstructed Siemens star from experimental data (a, b) and simulation data (c, d) for an initial angle of 68$^\circ$ where the angle was kept constant (a, c) or optimized in the model (b, d). Multiplication with the reference generates dark streaks wherever the reconstructed Siemens star spokes are 'out of phase' with the reference to produce a Moiré pattern.}
    \label{fig:EX0033_SI0007_siemens_star_ROI_2}
    \end{centering}
\end{figure}

The misalignment artifacts of the Siemens star indicate wherever the model is internally inconsistent, but this does not indicate likeness to the true image provided by the reference measurement. An absolute measure of reconstruction fidelity was however still possible by making use of the Moiré effect. When two Siemens star samples are overlayed, but one is displaced or distorted with respect to the other, a Moiré pattern forms due to the periodicity in the sample. We demonstrate this by multiplying the reconstruction pixel-by-pixel with the absolute value of the SEM reference after aligning the reference and reconstruction by maximizing their correlation. To prevent misalignment, the top and bottom rows of the Siemens star, which differed between the simulation and experimental reconstructions due to course alignment of the scanning positions, were disregarded for the correlation. The resulting image after multiplication can be interpreted as visualising the per-pixel contribution to the correlation coefficient between the reconstruction and the amplitude of the reference measurement: wherever a post-multiplication pixel is much darker than the original reconstruction its spokes are precisely a half cycle 'out of phase' with the reference, thus lowering the contribution to the total correlation coefficient at this point. A closeup of the entire region of interest consisting of just the Siemens star is shown in Fig. \ref{fig:EX0033_SI0007_siemens_star_ROI_2} for reconstructions performed on both experimental and simulation datasets. The full field of view for the experimental data can be found in the supplemental material.

For the constant tilt reconstructions, multiple vertical dark streaks across the Siemens star indicate that the reconstructions show poor fidelity with respect to the SEM reference. For the optimized case in simulation these dark streaks completely vanished, leaving only a small area in the center of the Siemens star out of phase. The optimized case on experimental data also completely removed the vertical streaks, leaving only some small areas near the high spatial frequencies in the center out of phase.

\medskip

\section{Conclusion}

We have demonstrated a novel approach to reflection ptychography, where the tilted propagation was included into the forward model through bilinear interpolation of the far-field coordinates. The automatic differentiation (AD)-based algorithm was tested using $450$ nm illumination for initial angles in the range of $70^\circ \pm 2^\circ$, where it showed convergence up to a precision of $\pm 0.05^\circ$. Furthermore, the approach was applied to data in the extreme ultraviolet (EUV) regime, where it resulted in both reduction of continuity artifacts and fidelity improvements when compared to a scanning electron microscope reference. By removing the need for manual estimation of the tilted-plane correction parameters, the new algorithm provides a more robust and reproducible approach to ptychographic reconstructions in a reflection geometry. The flexibility of AD-based modeling aligns with the broader progress in computational imaging by further shifting complexity from the instrumentation to the computational domain. AD-based modeling can effectively handle intricate imaging challenges that were previously out of reach and thus paves the way for future innovations in the field.

\begin{backmatter}
\bmsection{Funding}

TKI Holland High Tech (TKI HTSM/22.0220)\\
ASML Netherlands B.V.

\bmsection{Acknowledgments}
The authors wish to thank M. Dubbelman for his help in generating the white light interferometer measurements used in this publication.
\bmsection{Disclosures} 
SW: ASML Netherlands B.V. (E), CP: ASML Netherlands B.V. (E), WC: ASML Netherlands B.V. (E).











\bmsection{Data availability} Data underlying the results will be made available on the Coherent X-Ray Imaging Data bank (CXIDB).









\bmsection{Supplemental document}
See Supplement 1 for supporting content. 

\end{backmatter}

\section{References}

\bibliography{./references.bib, manual_references.bib}






\end{document}